\documentclass[sigconf, screen, authorversion]{acmart} %Arxiv version

\usepackage{amsfonts}
\usepackage{graphicx}
\usepackage{subcaption}
\usepackage{siunitx}
\usepackage{enumitem}
\usepackage{microtype}
\usepackage{textcomp}
\usepackage{appendix}
\usepackage{appendix}
\usepackage{pgfplots}
\pgfplotsset{compat=1.18} % Use a recent compatibility level
\usetikzlibrary{pgfplots.groupplots}
\usepgfplotslibrary{colorbrewer}

\definecolor{myblue}{HTML}{0072B2}  % A professional blue
\definecolor{myorange}{HTML}{E69F00} % A complementary orange

\pgfplotsset{
    custom-bar-chart/.style={
        width=0.90\textwidth,           % Adjust width
        height=0.50\textwidth,          % Adjust height
        ybar,                           % It's a bar chart
        bar width=10pt,                 % Adjust bar width for grouped bars
        enlarge x limits=0.2,           % Space at the ends of the x-axis
        ymin=0, ymax=7.5,               % Set y-limits, with a little headroom
        ylabel={Mean threat level (1--7)}, % Y-axis label
        ylabel style={font=\small},     % Make label font smaller
        symbolic x coords={Text, Image, Audio, Video}, % X-axis categories
        xtick=data,                     % Ticks at each data point
        tick label style={font=\small}, % Make tick labels smaller
        axis y line*=left,               % Only show left y-axis line
        axis x line*=bottom,             % Only show bottom x-axis line
        axis line style={thin, gray},   % Make axis lines thin and gray
        grid=major,                     % Add a major grid
        grid style={dashed, lightgray}, % Style the grid to be subtle
        nodes near coords,              % Display value on top of each bar
        nodes near coords style={
            font=\tiny,                 % Make the font tiny to avoid clutter
            yshift=8pt,                 % Shift them up to clear the error bar
            /pgf/number format/.cd,     % Format the numbers
                fixed,
                fixed zerofill,
                precision=1,
        },
        legend style={
            at={(0.5, 1.15)},           % Position legend above the plot
            anchor=north,
            legend columns=-1,          % Make legend horizontal
            draw=none,                  % No border around the legend
            fill=none,                  % No background color
            font=\small,
        },
    }
}

\usepgfplotslibrary{patchplots} 
\usepackage{hyperref}

\usepackage[hyphenbreaks]{breakurl}
\usepackage{xurl}
\def\BibTeX{{\rm B\kern-.05em{\sc i\kern-.025em b}\kern-.08em
    T\kern-.1667em\lower.7ex\hbox{E}\kern-.125emX}}
\usepackage{cleveref}
\usepackage{booktabs}
    
\usepackage{xcolor}

 \usepackage{pifont}% http://ctan.org/pkg/pifont

\usepackage{listings}
\usepackage{fancyvrb}
\usepackage{framed}
\usepackage{color}
\usepackage{pgfplots}
\usepackage{tikz}
\usetikzlibrary{positioning, decorations.pathreplacing, arrows.meta, shapes.geometric, shapes.misc, shapes.symbols, shapes.arrows, shapes.callouts, shapes.multipart, shapes.gates.logic.US, shapes.gates.logic.IEC, er, automata, backgrounds, petri, topaths, trees, arrows, matrix}

\usepackage{orcidlink}
\usepackage{tabularx}
\definecolor{light-gray}{gray}{0.95}
\hypersetup{
   breaklinks=true,   % splits links across lines
   colorlinks=true    % displays links as colored text instead of blocks
}

\usepackage{fancyhdr}

\copyrightyear{2026}
\acmYear{2026}
\setcopyright{cc}
\setcctype{by}
\acmConference[WWW Companion '26]{Companion Proceedings of the ACM Web Conference 2026}{April 13--17, 2026}{Dubai, United Arab Emirates}
\acmBooktitle{Companion Proceedings of the ACM Web Conference 2026 (WWW Companion '26), April 13--17, 2026, Dubai, United Arab Emirates}
\acmPrice{}
\acmDOI{10.1145/3774905.3795484}
\acmISBN{979-8-4007-2308-7/2026/04}
 
\begin{document}

\title[The Verification Crisis]{The Verification Crisis: Expert Perceptions of GenAI Disinformation and the Case for Reproducible Provenance}

\author{Alexander Loth}
\email{alexander.loth@stud.fra-uas.de}
\orcid{0009-0003-9327-6865}
\affiliation{%
  \institution{Frankfurt University of Applied Sciences}
  \city{Frankfurt am Main}
  \country{Germany}
}

\author{Martin Kappes}
\email{kappes@fra-uas.de}
\orcid{0000-0002-8768-8359}
\affiliation{%
  \institution{Frankfurt University of Applied Sciences}
  \city{Frankfurt am Main}
  \country{Germany}
}

\author{Marc-Oliver Pahl}
\email{marc-oliver.pahl@imt-atlantique.fr}
\orcid{0000-0001-5241-3809}
\affiliation{%
  \institution{IMT Atlantique, UMR IRISA, Chaire Cyber CNI}
  \city{Rennes}
  \country{France}
}

\begin{abstract}
The growth of Generative Artificial Intelligence (GenAI) has shifted disinformation production from manual fabrication to automated, large-scale manipulation. This article presents findings from the first wave of a longitudinal expert perception survey (N=21) involving AI researchers, policymakers, and disinformation specialists. It examines the perceived severity of multimodal threats---text, image, audio, and video---and evaluates current mitigation strategies.

Results indicate that while deepfake video presents immediate ``shock'' value, large-scale text generation poses a systemic risk of ``epistemic fragmentation'' and ``synthetic consensus,'' particularly in the political domain. The survey reveals skepticism about technical detection tools, with experts favoring provenance standards and regulatory frameworks despite implementation barriers.

%This report aligns with R2CASS (Reproducible and Reusable Computational Approaches) workshop objectives by arguing that 
GenAI disinformation research requires reproducible methods. 
The current challenge is measurement: without standardized benchmarks and reproducibility checklists,
%as advocated ~\cite{momeni2025checklists,bleier2025methodshub,gangopadhyay2025telegram}, 
tracking or countering synthetic media remains difficult. 
We propose treating information integrity as an infrastructure with rigor in data provenance and methodological reproducibility.
\end{abstract}

\begin{CCSXML}
<ccs2012>
 <concept>
  <concept_id>10010147.10010257.10010293.10010294</concept_id>
  <concept_desc>Computing methodologies~Artificial intelligence</concept_desc>
  <concept_significance>500</concept_significance>
 </concept>
 <concept>
  <concept_id>10002978.10003022.10003023</concept_id>
  <concept_desc>Security and privacy~Social aspects of security and privacy</concept_desc>
  <concept_significance>300</concept_significance>
 </concept>
</ccs2012>
\end{CCSXML}

\ccsdesc[500]{Computing methodologies~Artificial intelligence}
\ccsdesc[300]{Security and privacy~Social aspects of security and privacy}

\keywords{Verification crisis, reproducible provenance, GenAI disinformation, expert perception, epistemic fragmentation, synthetic consensus, computational reproducibility, deepfake detection, C2PA, Methods Hub}

\maketitle

%%% ========================================
%%% ARXIV PREPRINT HEADER (comment out for camera-ready)
\pagestyle{fancy}
\fancyhf{}
\fancyhead[L]{\textit{A.\ Loth, M.\ Kappes, M.-O.\ Pahl}}
\fancyhead[R]{\textit{Accepted at TheWebConf '26 Companion}}
\fancyfoot[C]{\thepage}
\renewcommand{\headrulewidth}{0pt}
%%% ========================================

\section{Introduction}
\label{sec:introduction}

The digital information ecosystem faces a structural shift in the economics of deception---a \emph{verification crisis}~\cite{wardle2017information,lazer2018science}. The maturation and accessibility of Large Language Models (LLMs) and diffusion-based image generation have reduced the marginal cost of producing high-fidelity disinformation toward zero~\cite{loth2024blessing,musser2023cost}. We observe a transition from ``disinformation as a craft''---characterized by labor-intensive troll farms, manual image manipulation, and human-operated botnets---to ``disinformation as an industry.'' This industrial mode enables automated, personalized, and persistent synthetic content at scale; LLMs can already simulate persona-specific public opinion~\cite{karanjai2025synthesizing}. Such a shift risks eroding the shared factual basis that underpins democratic deliberation, scientific discourse, and social cohesion---a phenomenon we term \emph{epistemic fragmentation}.

This paper examines this emerging threat through expert perspectives. By surveying AI engineers, policymakers---including members of the European Commission---and academic researchers ($n=21$), we map the GenAI threat landscape as perceived by practitioners. However, identifying the threat is insufficient. As our respondents observe, the traditional detection-versus-generation dynamic is failing~\cite{loth2026collateraleffects,loth2026eroding}. Generative capabilities advance faster than detection methods, and defensive tools remain opaque, unverified, and short-lived.

% In the context of the 2nd Workshop on Social Science Meets Web Data: Reproducible and Reusable Computational Approaches (R2CASS) at The ACM Web Conference 2026 (April 13--17, 2026, Dubai, United Arab Emirates), t
This paper argues that the solution lies in the rigorous application of reproducible science. The ``black box'' nature of commercial GenAI models, combined with the opacity of social media algorithms, creates a ``reproducibility crisis'' in disinformation studies~\cite{schoch2024reproducibility,loth2026collateraleffects}. If researchers cannot reproduce the generation of a threat or the performance of a detector, policy interventions become speculative and potentially harmful.

%Drawing on work from the R2CASS organizers, t
This paper integrates computational reproducibility checklists as proposed by Momeni, Khan, Kiesel, and Ross-Hellauer~\cite{momeni2025checklists}, reusable method repositories like the Methods Hub advocated by Bleier et al.~\cite{bleier2025methodshub}, and structured knowledge graphs for tracking claims as developed by Gangopadhyay, Dess{\`i}, and colleagues~\cite{gangopadhyay2024factchecking, gangopadhyay2025telegram, gangopadhyay2023claimskg}. The survey data indicate that experts favor \emph{reproducible provenance}---transparent, standardized infrastructure for verifying information origins---over opaque detection systems.

\subsection{Contributions: A Three-Act Narrative}
\label{subsec:contributions}

This paper tells the story of a \textbf{Crisis of Verification} through three acts:

\begin{enumerate}[leftmargin=*]
    \item \textbf{Act I: The Industrialization of Lying.} We examine how experts perceive the varying risks of text, audio, image, and video modalities. The data show that experts are concerned not only with \emph{content} but also with \emph{volume} and \emph{personalization}. AI-generated text (Mean 6.1/7) poses a systemic risk by polluting everyday discourse, while Deepfake Video (Mean 6.4/7) attracts attention but is more readily debunked.
    
    \item \textbf{Act II: The Failure of the Black Box.} We analyze why technical detection receives low confidence ratings (mean effectiveness $\approx 3.4/7$). Commercial detection systems are failing because they are \emph{black boxes}. Experts rate them as the least effective strategy. This reflects a failure of reproducibility: if we cannot verify how a detector works, we cannot trust its claims.
    
    \item \textbf{Act III: Reproducible Resistance}
    %The R2CASS Solution.} We connect expert feedback directly to the methodological innovations spearheaded by the R2CASS organizers. 
    To survive this crisis, we need ``Civil Engineering for Truth''---standardized, reproducible protocols. We introduce Momeni's Checklists~\cite{momeni2025checklists} and Bleier's Methods Hub~\cite{bleier2025methodshub} as the necessary infrastructure to build valid detectors and verify provenance.
\end{enumerate}

This paper argues for aligning computational social science with principles of rigor, reproducibility, and reusability. The ongoing expert survey is available at \url{https://github.com/aloth/verification-crisis}.

\section{Theoretical Framework: Reproducible Resistance and the Science of Truth}
\label{sec:framework}

To contextualize our survey findings, we situate them within the framework of computational reproducibility and web data analysis.
%central to the R2CASS workshop. The workshop aims to advance computational reproducibility in social science, particularly when analyzing digital behavioral data from platforms that are volatile and opaque.

\subsection{The Reproducibility Imperative in Social Data}

The analysis of digital behavioral data has historically been plagued by a lack of standardization. ``One-off'' scripts, inaccessible datasets, and undocumented parameters render much of the research in this field ephemeral. As noted by Bleier et al.~\cite{bleier2025methodshub} regarding the Methods Hub initiative, enabling reproducible and user-friendly computational tools is essential for the maturation of computational social science. The Methods Hub serves as a centralized repository where researchers can find open-source, quality-tested, and reproducible methods for data collection and analysis.

This infrastructure matters because our survey respondents call for ``society-wide infrastructure for provenance'' and note that ``coordination is a massive challenge.'' The fragmented nature of current detection tools---often proprietary or undocumented---prevents cumulative knowledge building. Schoch et al.'s work on defining computational reproducibility---a tier system based on who can reproduce a result (the original authors, trusted third parties, or everyone) and with which materials~\cite{schoch2024reproducibility}---provides the taxonomy needed to audit these tools. If a detection algorithm cannot be containerized and re-executed on a new dataset, it functions as a black box rather than a scientific instrument.

\subsection{Benchmarking and Checklists: The Momeni-Khan Framework}

Momeni, Khan, Kiesel, and Ross-Hellauer~\cite{momeni2025checklists} have established the necessity of checklists for computational reproducibility in social sciences. Their work addresses the ``validity gap'' in computational methods. This framework is critical when analyzing our survey results regarding technical detection tools. Experts rated these tools as unreliable (mean effectiveness $\approx 3.4/7$). A primary reason cited is the lack of standardized evaluation; detection claims are often made on static datasets that do not reflect the adversarial reality of the web.

Adopting the Momeni-Khan checklist approach for disinformation detection papers would ensure that claims of ``99\% detection accuracy'' are reproducible and robust against real-world shifts in generative models. These checklists require researchers to document the exact versions of libraries, the seeds of random number generators, and the provenance of training data. In the context of GenAI, where a model update can render a detector obsolete overnight, such rigor is not optional---it is the only way to maintain epistemic security.

\subsection{Knowledge Graphs and Longitudinal Tracking}

The work of Dess{\`i} and colleagues on knowledge graphs, the TeleScope dataset~\cite{gangopadhyay2025telegram}, and ClaimsKG~\cite{gangopadhyay2023claimskg} addresses the ``epistemic fragmentation'' identified by our survey respondents. One expert noted, ``The risk is not just `fake news' but interactive, personalized persuasion at scale.'' To counter this, we need structured ways to map narratives across platforms. The investigation of fact-checking characteristics and biases~\cite{gangopadhyay2024factchecking} and longitudinal analysis of Telegram discourse~\cite{gangopadhyay2025telegram} provide the architectural blueprint for the ``AI-assisted verification'' tools requested by respondents.

Furthermore, Khan's research on characterizing digital behavioral traces~\cite{khan2025twitter} and online computational reproducibility~\cite{khan2025valistad}, together with GESIS work on multilingual LLM-based fact-checking~\cite{saju2025llmfactchecking}, addresses the ``vernacular gap'' highlighted by survey respondents from the Global South. Recent work on Urdu fake news detection~\cite{nazar2025urdu} exemplifies the kind of language-specific, reproducible research that must be scaled to combat global disinformation. The focus on reproducibility of these NLP pipelines ensures that detection models are not just biased towards high-resource languages like English but are robust across linguistic contexts. We cannot fight dynamic AI agents with static blocklists; we need dynamic, interconnected knowledge bases and reproducible NLP pipelines.

\subsection{Paper Structure}
\label{subsec:structure}

The remainder of this paper is structured to follow our three-act narrative: Section~\ref{sec:methodology} describes our survey methodology. Section~\ref{sec:results} presents \textbf{Act I} (The Industrialization of Lying) and \textbf{Act II} (The Failure of the Black Box) through our survey findings on threat perceptions and mitigation skepticism. Section~\ref{sec:discussion} presents \textbf{Act III} (Reproducible Resistance), discussing how reproducible infrastructure can address the crisis. Section~\ref{sec:conclusion} synthesizes the argument and calls for action.
\section{Methodology}
\label{sec:methodology}

\subsection{Recruitment and Sampling}
The study employed a purposive snowball sampling strategy to recruit experts with direct experience in disinformation research, AI development, or digital policy. Initial recruitment seeds were identified through three channels: (1) \textbf{LinkedIn} searches for roles such as ``Trust and Safety Lead'' and ``AI Policy Director''; (2) \textbf{Google Scholar} identification of authors with highly-cited papers ($>100$ citations) on deepfakes and neural fake news (2023--2025); and (3) speakers at relevant \textbf{conference proceedings} (e.g., ICWSM, ACM CCS). This process yielded a final cohort of $N=21$ experts, including members of the European Commission, executives at AI laboratories, and senior academics. Data collection occurred between July and December 2025.

\begin{itemize}[leftmargin=*]
    \item \textbf{Sample Size:} $N=21$ senior experts. Given the purposive sampling of high-level decision-makers (including a Member of the European Commission and executives at major AI companies), this sample prioritizes depth of insight over statistical power.
    \item \textbf{Respondent Verification:} All participants confirmed they were 18+ and consented to the study.
    \item \textbf{Geographic Distribution:} The sample is predominantly European, reflecting the region's leadership in digital regulation (e.g., the AI Act):
    \begin{itemize}
        \item European Union (EU): 52.4\%
        \item North America (USA \& Canada): 23.8\%
        \item Europe (non-EU): 14.3\%
        \item South \& Central Asia: 4.8\%
        \item East \& Southeast Asia: 4.8\%
    \end{itemize}
\end{itemize}

\subsection{Survey Instrument Design}
The survey, hosted on Google Forms (\url{https://forms.gle/BCwYFtfqxmZewkL97}), was designed to measure two latent constructs: \textit{Threat Severity} and \textit{Mitigation Efficacy}.

\subsubsection{Threat Severity Scale}
Respondents rated four modalities (Text, Image, Audio, Video) on a 7-point Likert scale.
\begin{itemize}
    \item \textbf{Anchor 1 (Not a threat):} ``The modality poses no significant risk beyond pre-GenAI baselines; current defenses are adequate.''
    \item \textbf{Anchor 7 (Severe threat):} ``The modality poses a severe risk to information integrity, with potential for widespread epistemic harm.''
\end{itemize}

\subsubsection{Mitigation Efficacy Scale}
Respondents evaluated five strategies (Literacy, C2PA, Regulation, Platforms, Detection).
\begin{itemize}
    \item \textbf{Anchor 1 (Not effective):} ``The strategy is easily bypassed, technically unfeasible, or counter-productive.''
    \item \textbf{Anchor 7 (Very effective):} ``The strategy provides a robust, scalable, and enduring defense.''
\end{itemize}

\subsection{Participant Demographics}

The respondent pool includes professionals with backgrounds in both technical and policy domains.

\begin{table*}[htbp]
\caption{Participant Professional Roles and Experience}
\label{tab:demographics}
\begin{tabularx}{\textwidth}{lrX}
\toprule
\textbf{Role Category} & \textbf{Count} & \textbf{Representative Titles} \\
\midrule
AI Researchers / Engineers & 9 (42.9\%) & ``VP @ social media company'', ``Professor Cybersecurity'', ``AI researcher'' \\
Policymakers / Regulators & 2 (9.5\%) & ``Member of the European Commission'', ``Architect of DPI frameworks'' \\
Journalists / Civil Society & 3 (14.3\%) & ``Spiegel-Kolumnist'', ``Disinformation/fact-checking professional'' \\
Industry Executives & 3 (14.3\%) & ``Exec of a leading AI company'', ``Partner at an AI fund'' \\
Other & 4 (19.0\%) & ``Teacher'', ``IT Architect'', ``Ethicist/Legal Scholar'' \\
\bottomrule
\end{tabularx}
\end{table*}

\paragraph{Experience Levels.} Respondents reported substantial professional experience:
\begin{itemize}[leftmargin=*]
    \item 35 Years: Technologist / Web Standards Advocate
    \item 28 Years: Member of the European Commission
    \item 25 Years: Journalist; Investor/VC
    \item \textbf{Median Experience:} $\approx$15 Years
\end{itemize}

\paragraph{Technical Literacy.}
\begin{itemize}[leftmargin=*]
    \item \textbf{LLM Understanding:} 90.5\% rated their understanding of Large Language Models as 4 or higher (out of 7). 28.6\% rated it 7 out of 7.
    \item \textbf{Deepfake Familiarity:} 42.9\% rated their familiarity with deepfake technology 7 out of 7.
\end{itemize}

\subsection{Qualitative Analysis Protocol}
Qualitative responses were analyzed using \textit{Reflexive Thematic Analysis} following the six-phase framework established by Braun and Clarke (2006; 2021) \cite{braun2006thematic, braun2021reflexive}.
\begin{enumerate}
    \item \textbf{Inductive Coding:} We adopted a data-driven approach, avoiding pre-existing codebooks to allow GenAI-specific concepts (e.g., ``Weaponized Intimacy'') to emerge from the data.
    \item \textbf{Theme Generation:} Initial codes such as \texttt{[volume]}, \texttt{[personalized persuasion]}, and \texttt{[consensus fabrication]} were grouped under the higher-order theme of ``Synthetic Consensus.'' Similarly, codes regarding \texttt{[detection failure]} and \texttt{[arms race]} were grouped under ``The Failure of the Black Box.''
    \item \textbf{Refinement:} Themes were reviewed against the quantitative data to ensure consistency. For instance, the qualitative theme of ``provenance as infrastructure'' corresponds to the high quantitative rating for C2PA.
\end{enumerate}

\subsection{Data Analysis Method}

Quantitative data regarding threat levels (1--7 scale) and mitigation effectiveness (1--7 scale) were analyzed using descriptive statistics to identify central tendencies and variance. Given the $N=21$ sample size, statistical significance testing was deprioritized in favor of identifying strong signals and consensus clusters.

Qualitative responses from open-ended questions were coded thematically using an inductive approach. We identified recurring concepts such as ``trust erosion,'' ``provenance,'' ``institutional failure,'' and ``epistemic fragmentation.'' These qualitative insights were then mapped against the quantitative data to explain why certain trends appeared (e.g., why detection is rated low despite high technical literacy).

\paragraph{Reproducibility Statement.} In adherence to R2CASS principles, the anonymized CSV dataset, the full survey instrument, and the Python/Pandas analysis code used to generate the statistics in this paper will be made available via the GESIS Methods Hub~\cite{bleier2025methodshub}. The reproducibility package follows the checklist guidelines established by Momeni et al.~\cite{momeni2025checklists}, documenting exact library versions, random seeds, and data provenance. This work builds on the JudgeGPT / RogueGPT research projects\footnote{\url{https://github.com/aloth/JudgeGPT}}\textsuperscript{,}\footnote{\url{https://github.com/aloth/RogueGPT}}, which provide open-source infrastructure for studying human perception of AI-generated content.

%======================================================================
% results.tex  (wide-table, two-column layout)
%======================================================================

\section{Results: The Threat Landscape}\label{sec:results}

The survey results indicate a threat landscape evolving faster than current defensive capabilities. Experts view GenAI disinformation not as a future risk but as a present problem.

%----------------------------------------------------------------------
\subsection{Modalities of Deception: The Hierarchy of Harm}\label{sec:threat}
%----------------------------------------------------------------------

The survey asked experts to rate the threat level of four GenAI modalities: Text, Images, Audio, and Video. The results show a high baseline of concern across all modalities, but with nuanced distinctions in how they threaten society.

\begin{table*}[htbp]
\caption{Perceived Threat Level by Modality (Mean Rating / 7)}
\label{tab:modality_threat}
\begin{tabularx}{\textwidth}{lrX}
\toprule
\textbf{Modality} & \textbf{Mean} & \textbf{Key Expert Insight} \\
\midrule
Deepfake Video & 6.4 & ``The ultimate tool for eroding the `shared objective reality' required for democracy\ldots It creates an environment of total doubt.'' \\
AI-Generated Text & 6.1 & ``Large-scale text generation is the most significant threat to eroding general veracity discernment\ldots The marginal cost\ldots is trending toward zero.'' \\
Voice Cloning (Audio) & 5.9 & ``Audio cloning presents an immediate, hard-to-detect vector for fraud\ldots Voice-based payment scams are the new phishing.'' \\
AI-Generated Images & 5.8 & ``It helps calibrate public expectations. People need to understand that they can be fooled.'' \\
\bottomrule
\end{tabularx}
\end{table*}

\paragraph{Analysis.} Deepfake Video is consistently rated highest (Mean 6.4) due to its ``shock value'' and the persuasive power of visual evidence; experimental evidence indicates that political deepfakes mainly create uncertainty, which in turn reduces trust in news~\cite{vaccari2020deepfakes}. However, the qualitative data reveals a deeper anxiety regarding Text (Mean 6.1). While video is episodic (a specific deepfake goes viral), text is \emph{systemic}. Text is the substrate of the internet---comments, reviews, news articles, tweets. Empirical studies on human perception confirm that LLM-generated text achieves near-human mimicry scores, making detection by readers unreliable~\cite{loth2026eroding,loth2026collateraleffects}.

Experts warn that GenAI text generation allows for ``Synthetic Consensus'' or ``Astroturfing'' at an industrial scale. Prior research has shown that false news spreads faster and farther than true news online~\cite{vosoughi2018spread}. One respondent, a Professor of Cybersecurity, noted: \emph{``Large-scale text spambots (astroturfing)\ldots have the most impact to our current information society.''} Another expert, an AI researcher, emphasized: \emph{``Personalized 1:1 persuasion via text is a bigger threat to democracy than a fake video which can be debunked.''} This distinction is critical: Video attacks the truth of an event, but text attacks the perception of public opinion. Research on the disinformation kill chain demonstrates a ``perception-accuracy gap''---suspicion does not improve detection accuracy~\cite{loth2026killchaingi}.

This aligns with Khan et al.'s work on digital behavioral data~\cite{khan2025twitter}. If the volume of synthetic text overwhelms organic human text, traditional social science methods that rely on scraping web data will fail. They will effectively be studying the output of LLMs rather than human behavior. This underscores the R2CASS mission: we need reproducible methods to filter and validate text data before it enters the social science pipeline.

%----------------------------------------------------------------------
\subsection{Domain-Specific Threats: A Comparative Analysis}\label{sec:domain}
%----------------------------------------------------------------------

The survey segmented threats across Political, Health, Financial, and Social domains. While political threats are the most visible, other domains face significant risks that receive less attention.

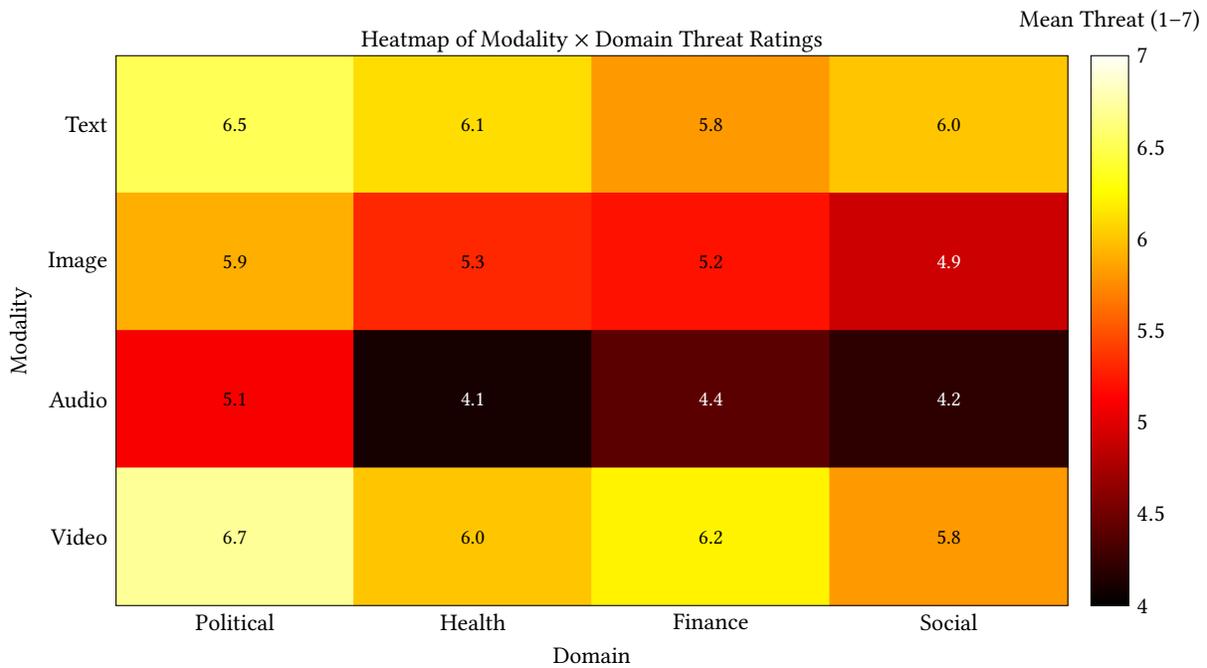
\begin{figure*}[htbp]
  \centering
  \begin{tikzpicture}
    \begin{axis}[
        % General Styling
        width=0.80\textwidth, height=0.50\textwidth,
        title={Heatmap of Modality $\times$ Domain Threat Ratings},
        title style={font=\normalsize, yshift=-8pt},
        xlabel={Domain}, ylabel={Modality},
        y dir=reverse,
        % Colormap and Colorbar
        colormap/hot2,
        colorbar,
        colorbar style={title={Mean Threat (1--7)}},
        point meta min=4, point meta max=7,
        % Axis Ticks
        xtick={0,1,2,3},
        xticklabels={Political,Health,Finance,Social},
        ytick={0,1,2,3},
        yticklabels={Text,Image,Audio,Video},
        tick style={draw=none},
        enlargelimits=false,
        axis on top,
      ]
      \addplot[matrix plot*, point meta=explicit, mesh/cols=4] coordinates {
        (0,0) [6.5] (1,0) [6.1] (2,0) [5.8] (3,0) [6.0]
        (0,1) [5.9] (1,1) [5.3] (2,1) [5.2] (3,1) [4.9]
        (0,2) [5.1] (1,2) [4.1] (2,2) [4.4] (3,2) [4.2]
        (0,3) [6.7] (1,3) [6.0] (2,3) [6.2] (3,3) [5.8]
      };
      % Manual labels with appropriate colors
      % Row 0 (Text): all >= 5.0, so black
      \node[font=\small] at (axis cs:0,0) {6.5};
      \node[font=\small] at (axis cs:1,0) {6.1};
      \node[font=\small] at (axis cs:2,0) {5.8};
      \node[font=\small] at (axis cs:3,0) {6.0};
      % Row 1 (Image): all >= 5.0 except 4.9
      \node[font=\small] at (axis cs:0,1) {5.9};
      \node[font=\small] at (axis cs:1,1) {5.3};
      \node[font=\small] at (axis cs:2,1) {5.2};
      \node[font=\small, white] at (axis cs:3,1) {4.9};
      % Row 2 (Audio): 5.1 black, rest white (< 5.0)
      \node[font=\small] at (axis cs:0,2) {5.1};
      \node[font=\small, white] at (axis cs:1,2) {4.1};
      \node[font=\small, white] at (axis cs:2,2) {4.4};
      \node[font=\small, white] at (axis cs:3,2) {4.2};
      % Row 3 (Video): all >= 5.0, so black
      \node[font=\small] at (axis cs:0,3) {6.7};
      \node[font=\small] at (axis cs:1,3) {6.0};
      \node[font=\small] at (axis cs:2,3) {6.2};
      \node[font=\small] at (axis cs:3,3) {5.8};
    \end{axis}
  \end{tikzpicture}
  \Description{Heatmap showing mean expert threat ratings (1--7) across four modalities (text, image, audio, video) and four domains (political, health, finance, social), with higher values concentrated in the political domain.}
  \caption{Threat Perception Heatmap. Heatmap of expert threat ratings showing the concentration of high-severity concerns in the Political domain compared to the Social domain. Note the skew towards ``Severe Threat'' (7) for Political Text and Video.}
  \label{fig:modality_domain}
\end{figure*}

\subsubsection{The Political Domain: The Immediate Emergency}

Respondents identified this domain as the primary vulnerability. 90.5\% selected ``Election interference via deepfake video'' as a top-3 urgent risk.

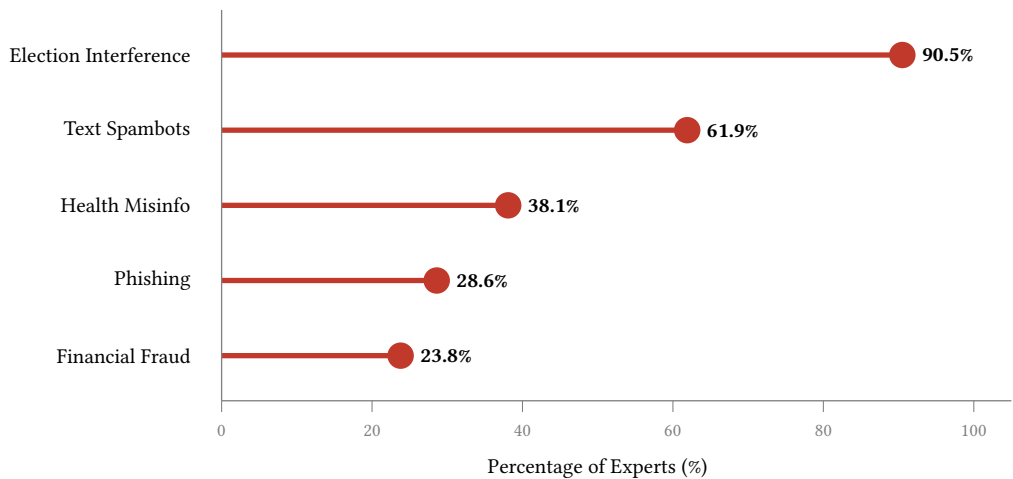
\begin{figure*}[htbp]
\centering
\begin{tikzpicture}
  % Define colors
  \definecolor{accent}{RGB}{192,57,43}
  
  % Y positions for each category
  \foreach \y/\val/\label in {
    5/90.5/Election Interference,
    4/61.9/Text Spambots,
    3/38.1/Health Misinfo,
    2/28.6/Phishing,
    1/23.8/Financial Fraud
  } {
    % Category label on left
    \node[anchor=east, font=\small] at (-0.3, \y) {\label};
    % Horizontal line (stem)
    \draw[accent, line width=2pt] (0, \y) -- (\val/10, \y);
    % Circle at end (lollipop head)
    \fill[accent] (\val/10, \y) circle (5pt);
    % Value label
    \node[anchor=west, font=\small\bfseries] at (\val/10 + 0.15, \y) {\val\%};
  }
  
  % X-axis
  \draw[gray, line width=0.5pt] (0, 0.4) -- (0, 5.6);
  \draw[gray, line width=0.5pt] (0, 0.4) -- (10.5, 0.4);
  
  % X-axis ticks and labels
  \foreach \x in {0, 2, 4, 6, 8, 10} {
    \draw[gray] (\x, 0.4) -- (\x, 0.25);
    \node[font=\footnotesize, gray] at (\x, 0) {\pgfmathparse{int(\x*10)}\pgfmathresult};
  }
  
  % X-axis label
  \node[font=\small] at (5, -0.5) {Percentage of Experts (\%)};
  
\end{tikzpicture}
\Description{Lollipop chart showing the percentage of experts selecting each item as a top-3 urgent risk; election interference ranks highest, followed by text spambots, health misinformation, phishing, and financial fraud.}
\caption{Urgency Ranking. Top-3 Urgent Risks selected by experts, highlighting Election Interference (90.5\%) and Text Spambots (61.9\%) as the dominant concerns, far outpacing Phishing or Financial Fraud.}
\label{fig:urgency_ranking}
\end{figure*}

\begin{quote}
\emph{``Democracy relies on a shared understanding of reality. When AI is used to manufacture consent or distort the truth during elections, it attacks the sovereign will of the people.''} \hfill (Member of the European Commission)
\end{quote}

\paragraph{Interpretation.} The experts differentiate between ``content'' and ``reaction.'' One respondent (VP at a social media company) noted, \emph{``The threat isn't just the content, it's the reaction to it\ldots Election interference is a societal problem where our role is to give people context, not to censor speech.''} This suggests that technical detection alone is insufficient; reproducible social science approaches to measuring impact---building on work on computational reproducibility in CSS by Schoch, Bleier, and colleagues~\cite{schoch2024reproducibility}---are required to understand how disinformation changes voting behavior.

\subsubsection{The Financial Domain: The Rise of Audio Fraud}

While political threats are about persuasion, financial threats are about extraction. The rise of audio-cloning is a specific, high-velocity threat here.

\begin{quote}
\emph{``Voice-based payment scams are the new phishing.''} \hfill (Cybersecurity Expert)
\end{quote}

\begin{quote}
\emph{``Crypto and finance rely on cryptographic truth. AI attacks that foundation.''} \hfill (FinTech Executive)
\end{quote}

\paragraph{Interpretation.} The ``crypto-provenance'' solution mentioned by a respondent highlights a potential intersection between Web3 technologies and R2CASS's focus on data integrity. If financial markets cannot trust the voice on the phone, the friction in the economy increases drastically.

\subsubsection{The Social Domain: Weaponized Intimacy}

A notable finding is the emergence of ``Weaponized Intimacy'' or the ``Psychosis Risk.''

\begin{quote}
\emph{``Looking ahead five years, a major, currently underestimated risk is `Weaponized Intimacy' (Persuasive AI Companions). While we currently focus on deepfakes (fake content), the next frontier is deep agents (fake relationships).''} \hfill (AI Ethics Researcher)
\end{quote}

\begin{quote}
\emph{``The `Psychosis Risk.' That millions of people will fall into deep, emotionally dependent relationships with AI companions that are manipulated by bad actors. Disinformation won't just be a news article; it will come from your `best friend' (the AI).''} \hfill (Technology Executive)
\end{quote}

\paragraph{Implication.} This represents a shift from broadcasting lies to micro-targeting emotional manipulation. Studying this requires new, reproducible methodologies. Dess{\`i}'s TeleScope dataset~\cite{gangopadhyay2025telegram}, which looks at Telegram channels, is a perfect example of the kind of data needed. Telegram is often used for more intimate, community-based communication than Twitter/X. To study ``deep agents,'' researchers will need longitudinal datasets of interaction, not just static snapshots of posts.

%----------------------------------------------------------------------
\subsection{The ``Epistemic Fragmentation''}\label{sec:fragmentation}
%----------------------------------------------------------------------

A recurring theme in the qualitative data is the fear that GenAI will destroy the possibility of knowing anything---a state of \emph{Epistemic Fragmentation}.

\begin{quote}
\emph{``The risk is not that people believe a specific lie, but that they stop believing in the possibility of truth altogether.''} \hfill (Policy Advisor)
\end{quote}

\begin{quote}
\emph{``Epistemic fragmentation is more dangerous than any single fake video.''} \hfill (AI Researcher)
\end{quote}

\begin{quote}
\emph{``The `Dead Internet Theory' is becoming reality. When 99\% of content is synthetic, the remaining humans will retreat into smaller, closed tribes to find truth.''} \hfill (Journalist)
\end{quote}

This ``Reality Apathy'' poses a challenge for the R2CASS community: If the ``Web Data'' we study is largely synthetic, our social science becomes a study of bot behavior. Bleier's work on computational reproducibility~\cite{bleier2025methodshub} becomes the defensive wall against this, while structured approaches like ClaimsKG~\cite{gangopadhyay2023claimskg} provide the infrastructure for tracking claim truthfulness at scale. We must be able to prove that our datasets are human, or at least understand the synthetic distribution within them.

%======================================================================
\subsection{Mitigation Strategies: The Failure of Detection}\label{sec:mitigation}
%======================================================================

The survey asked experts to rate and rank mitigation strategies. The results show low confidence in purely technical solutions, particularly detection tools.

\begin{table*}[htbp]
\caption{Effectiveness Ratings of Mitigation Strategies (1--7 Scale)}
\label{tab:mitigation}
\begin{tabularx}{\textwidth}{lrX}
\toprule
\textbf{Strategy} & \textbf{Mean} & \textbf{Primary Limitation Cited} \\
\midrule
Media / Public Literacy & 4.8 & ``Reaches only tech-savvy audiences''; ``Risks increasing general distrust''; ``Impact fades once novelty wears off.'' \\
Digital Watermarking / C2PA & 4.2 & ``Easily bypassed by adversarial content'' (e.g., stripping metadata); ``Coordination\ldots massive challenge.'' \\
Government Regulation & 4.1 & ``Pacing problem'' (Laws are too slow); ``Stifles open-source innovation''; ``Europe is regulating itself out of relevance.'' \\
Platform Enforcement & 3.5 & Business model conflict (``Platforms profit from engagement, not truth''). \\
Technical Detection Tools & 3.4 & ``The `cat and mouse' dynamic\ldots detection becomes statistically harder''; ``Unreliable.'' \\
\bottomrule
\end{tabularx}
\end{table*}

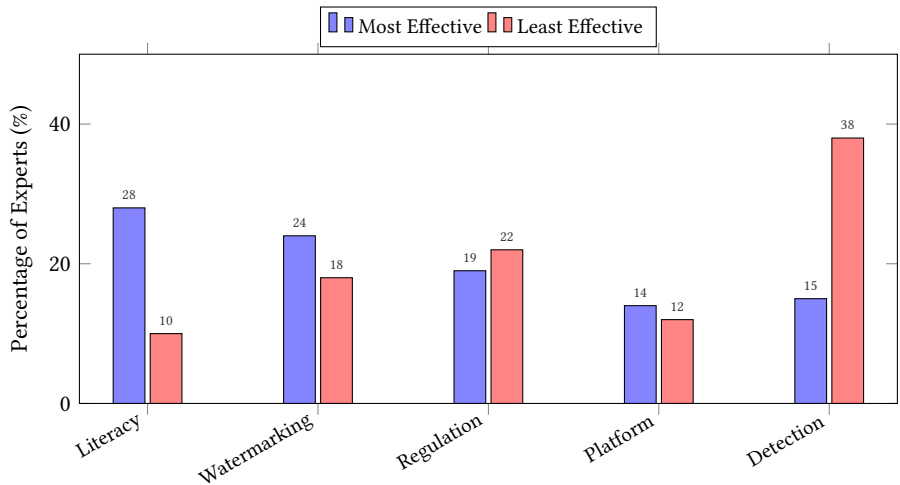
\begin{figure*}[htbp]
\centering
\begin{tikzpicture}
  \begin{axis}[
    ybar,
    bar width=12pt,
    width=0.70\textwidth,
    height=0.35\textwidth,
    ylabel={Percentage of Experts (\%)},
    symbolic x coords={Literacy, Watermarking, Regulation, Platform, Detection},
    xtick=data,
    x tick label style={rotate=30, anchor=east, font=\small},
    ymin=0, ymax=50,
    legend style={at={(0.5,1.02)}, anchor=south, legend columns=2, font=\small},
    nodes near coords,
    nodes near coords style={font=\tiny},
    every axis plot/.append style={fill opacity=0.8},
  ]
  \addplot[fill=blue!60] coordinates {(Literacy, 28) (Watermarking, 24) (Regulation, 19) (Platform, 14) (Detection, 15)};
  \addplot[fill=red!60] coordinates {(Literacy, 10) (Watermarking, 18) (Regulation, 22) (Platform, 12) (Detection, 38)};
  \legend{Most Effective, Least Effective}
  \end{axis}
\end{tikzpicture}
\Description{Grouped bar chart comparing the percentages of experts who rated each mitigation strategy as most effective versus least effective; technical detection is most often rated least effective.}
\caption{The Mitigation Gap. Best-Worst Scaling results indicating that Technical Detection Tools are viewed as the least effective strategy (38\% rated ``Least Effective''), contradicting the current industry focus on ``AI catchers.'' Provenance and Literacy are favored but face implementation hurdles.}
\label{fig:bws_mitigation}
\end{figure*}

%----------------------------------------------------------------------
\subsubsection{The Detection Paradox: Why ``AI Catching AI'' is Failing}\label{sec:detection_paradox}
%----------------------------------------------------------------------

Experts view Technical Detection Tools as the least effective long-term solution (Mean 3.4). Prior work on neural fake news generation and detection has documented the inherent difficulty of this task~\cite{zellers2019grover}. A respondent explained:

\begin{quote}
\emph{``As models improve, detection becomes statistically harder\ldots The `cat and mouse' dynamic.''} \hfill (AI Researcher)
\end{quote}

Another noted:

\begin{quote}
\emph{``Ultimately, we need AI to detect AI. Humans won't keep up. But detection is unreliable\ldots it evokes the impression of reliability.''} \hfill (Cybersecurity Expert)
\end{quote}

\paragraph{R2CASS Relevance.} This finding underscores the need for reproducible evaluation. Current commercial detectors are often ``black boxes'' with unpublished accuracy rates on novel datasets. When a new model (e.g., GPT-5) is released, existing detectors often fail. Moreover, recent work shows that adversarial perturbations can substantially degrade AI-generated text detection, making many detectors brittle against a determined adversary~\cite{khalid2025adversarial}. The community needs the checklists proposed by Momeni et al.~\cite{momeni2025checklists} to audit these tools.

\paragraph{The Checklist Solution.} Momeni's framework requires detailing the training data, the preprocessing steps, and the evaluation metrics. If a detection paper claims high accuracy but does not provide these details (as is common), it contributes to the ``illusion of reliability'' that the experts fear. This is especially critical because detector performance can drop sharply when moving across datasets, generators, and post-processing pipelines, undermining real-world generalization~\cite{kose2025certainly}. R2CASS can lead the way by mandating these checklists for all submissions.

%----------------------------------------------------------------------
\subsubsection{The Provenance (C2PA) Hope \& Skepticism}\label{sec:provenance}
%----------------------------------------------------------------------

While some experts see C2PA (Coalition for Content Provenance and Authenticity) as ``essential infrastructure,'' others note it creates a ``two-tier internet.'' Tools like Origin Lens~\cite{loth2026originlens} demonstrate that cryptographic verification of Content Credentials can be implemented on consumer devices, but adoption remains fragmented.

\begin{quote}
\emph{``Watermarking only works if everyone uses it. If `rogue' open-source models do not implement these safety standards, the ecosystem remains polluted.''} \hfill (Policy Advisor)
\end{quote}

\paragraph{Tension.} This highlights a tension between Open Science (which R2CASS supports) and AI Safety. Open weights allow for reproducibility, but also for the removal of watermarks. Recent advances in LLM watermarking~\cite{kirchner2023watermark} offer promise, but face similar adoption challenges. The tier system of computational reproducibility by Schoch et al.~\cite{schoch2024reproducibility}, which distinguishes materials shared only with trusted third parties from fully public ones, suggests a middle ground: such ``Tiered Reproducibility'' could be applied to safety mechanisms.

%----------------------------------------------------------------------
\subsubsection{Literacy as an ``Immune System''}\label{sec:literacy}
%----------------------------------------------------------------------

There is a split on media literacy. Some view it as the only viable defense:

\begin{quote}
\emph{``Literacy is the `user interface' for truth.''} \hfill (Journalist)
\end{quote}

Others warn of ``Skepticism Spillover'':

\begin{quote}
\emph{``It risks increasing general distrust\ldots making people skeptical of all content.''} \hfill (AI Ethics Researcher)
\end{quote}

\paragraph{Recommendation.} Literacy tools must be rigorous. Research on the psychology of fake news suggests that analytical thinking can reduce susceptibility to misinformation~\cite{pennycook2021psychology}. The mention of ``gamification'' and ``inoculation'' suggests a need for reproducible social science experiments to test which literacy interventions actually work. We cannot rely on intuition; we need reproducible studies, potentially hosted on the Methods Hub~\cite{bleier2025methodshub}, to share effective literacy curricula.

%======================================================================

%======================================================================
\section{Discussion: Toward a Reproducible Resistance}\label{sec:discussion}
%======================================================================

The survey data indicate a mismatch between threat scale and defense capacity. The threats are industrial, personalized, and multimodal, while defenses remain fragmented and easily bypassed. Experts call for structural change.

%----------------------------------------------------------------------
\subsection{The Need for a ``Methods Hub'' for Disinformation}\label{sec:methodshub_need}
%----------------------------------------------------------------------

The varying perceptions of threat and mitigation effectiveness stem from a lack of shared data and methods. One expert noted:

\begin{quote}
\emph{``We need a global scientific agreement that certain safety features are mandatory.''} \hfill (Technology Executive)
\end{quote}

This is exactly the mission of the GESIS Methods Hub~\cite{bleier2025methodshub}. To address the ``epistemic fragmentation,'' we need a centralized, open-access repository where:

\begin{enumerate}[leftmargin=*]
    \item \textbf{Datasets of synthetic vs.\ organic content are curated and versioned.} The TeleScope dataset by Gangopadhyay, Dess{\`i}, Dimitrov, and Dietze~\cite{gangopadhyay2025telegram} is a prime example of a curated, citable corpus of organic platform data that such efforts can build on. It provides a longitudinal view of Telegram, a platform rife with disinformation. By making this dataset available and citable, the authors allow the entire community to test detection hypotheses on real-world data, not just toy datasets.
    
    \item \textbf{Detection Algorithms are hosted in reproducible containers} (Docker/Binder), allowing independent verification of their efficacy---addressing the ``unreliable detection'' concern raised by respondents.
    
    \item \textbf{Reproducibility Checklists}~\cite{momeni2025checklists} are applied to all new findings on AI influence, ensuring that policy is based on solid science, not hype.
\end{enumerate}

%----------------------------------------------------------------------
\subsection{Narrative Shift: From ``Truth'' to ``Provenance''}\label{sec:provenance_shift}
%----------------------------------------------------------------------

Our findings suggest pivoting the narrative from the philosophical (and computationally intractable) goal of ``Determining Truth'' to the technical and reproducible goal of ``Verifying Provenance.'' The survey respondents repeatedly emphasized that they do not want AI to be the ``arbiter of truth.'' Instead, they want tools that show \emph{context}: Where did this come from? Is it human?

\paragraph{Proposed Narrative Arc.}

\begin{enumerate}[leftmargin=*]
    \item \textbf{The Diagnosis:} The survey confirms that GenAI has weaponized the ``Information Gap'' (asymmetry between generation cost and verification cost).
    
    \item \textbf{The Failure:} Current ad-hoc detection fails because it is not robust or reproducible. It creates a false sense of security. Empirical data show that human fake-news detection accuracy degrades under sustained exposure (by 10.2 percentage points), while AI-origin detection remains stable~\cite{loth2026killchaingi}, consistent with what our broader research agenda terms the indistinguishability threshold, the point at which human detection accuracy converges to chance~\cite{loth2026symposium}.
    
    \item \textbf{The Solution:} We must build ``Epistemic Infrastructure'' based on open standards. This involves integrating Knowledge Graphs~\cite{gangopadhyay2023claimskg} to track narrative evolution, Reproducible Workflows~\cite{khan2025valistad,bleier2025methodshub} to audit defenses, and reproducible human perception evaluation pipelines~\cite{loth2026collateraleffects}.
\end{enumerate}

%----------------------------------------------------------------------
\subsection{Operationalizing Reproducible Provenance}\label{sec:operationalizing}
%----------------------------------------------------------------------

The expert consensus points away from \textit{post-hoc} detection and toward \textit{pre-hoc} provenance. However, to transition ``provenance'' from a theoretical ideal to a deployable infrastructure requires concrete metrics. We propose a framework integrating the C2PA technical standard with the reproducibility requirements of the Methods Hub.

\subsubsection{Technical Metrics: The C2PA Standard}
The Coalition for Content Provenance and Authenticity (C2PA) provides a cryptographic foundation for content authentication~\cite{c2pa2025spec,loth2026originlens}. To evaluate the readiness of this infrastructure, we propose the following metrics:

\begin{itemize}
    \item \textbf{Manifest Robustness:} Defined as the survival rate of provenance assertions (signatures) after standard social media transcoding. Operational goal: $>95\%$ survival of the \texttt{c2pa} box in ISO BMFF containers.
    \item \textbf{Verification Latency:} The computational overhead introduced by signature validation at the client side. To prevent user friction, we propose targets of $<100$ms for static images and $<500$ms for video streams.
    \item \textbf{Chain Continuity:} The percentage of an asset's lifecycle (Capture $\rightarrow$ Edit $\rightarrow$ Publish) covered by unbroken cryptographic assertions.
\end{itemize}

\subsubsection{Methodological Metrics: The TOM Score}
Provenance applies not only to media but to the scientific tools used to study it. To address reproducibility concerns in disinformation detection, we propose a \textit{Transparency of Methods (TOM)} score, defined as the share of applicable items of the computational reproducibility checklists by Momeni et al.~\cite{momeni2025checklists} (as used on the Methods Hub) that a tool fulfills. We suggest that AI-based detection tools used in policy decisions should target a TOM score $>0.8$, documenting:
\begin{enumerate}
    \item \textbf{Data Provenance:} Exact composition of Synthetic vs. Organic training data.
    \item \textbf{Model Genealogy:} Full specification of the pre-trained base models (e.g., BERT, ViT) and fine-tuning hyperparameters.
    \item \textbf{Adversarial Robustness:} Performance metrics against known adversarial attacks, not just static benchmarks.
\end{enumerate}

Adopting these metrics shifts the discourse from abstract notions of ``trust'' toward measurable, auditable standards.

%----------------------------------------------------------------------
\subsection{Future Risks: The ``Psychosis Risk'' and Human-AI Forensics}\label{sec:future_risks}
%----------------------------------------------------------------------

An emerging risk identified in the survey is \emph{Anthropomorphism} or the ``Psychosis Risk''; notably, empirical evidence suggests AI companions can reduce loneliness, which may help explain why some users become emotionally dependent on them~\cite{defreitas2025ai}:

\begin{quote}
\emph{``Millions of people will fall into deep, emotionally dependent relationships with AI companions\ldots Disinformation\ldots will come from your `best friend'.''} \hfill (Technology Executive)
\end{quote}

This requires a new branch of computational social science: \textbf{Human-AI Interaction Forensics}. We need to develop reproducible methods to study how users bond with agents and how those bonds are exploited. This links back to work on validating social text-as-data methods and online computational reproducibility~\cite{khan2025valistad}---we need to analyze the semantics of these human-AI conversations to detect manipulative patterns.

The challenge is considerable: How do we study relationships that are designed to be private and emotionally intimate? Traditional social science methods (surveys, experiments) may be insufficient. We may need new computational approaches that can analyze interaction logs at scale while preserving privacy---a perfect use case for the reproducible, containerized methods that R2CASS promotes.

%======================================================================

%======================================================================
\section{Conclusion}\label{sec:conclusion}
%======================================================================

This Work-In-Progress report 
%for the R2CASS workshop 
documents expert perceptions of the emerging \emph{verification crisis}: the era of ``trusting your eyes'' is ending, and verifying sources must take precedence. The survey results ($N=21$) indicate that GenAI disinformation poses a multimodal threat to political and social discourse, driven by low production costs and detection difficulty. Experts identified ``Epistemic Fragmentation'' and ``Synthetic Consensus'' as primary concerns, with synthetic content risking displacement of human discourse.

However, the path forward is not simply ``better AI models.'' It is \emph{better science}. The skepticism regarding current mitigation strategies---particularly technical detection---underscores the need for reproducible and reusable computational approaches.
%as promoted by the R2CASS community. 
We cannot regulate or detect what we cannot consistently measure.

By adopting 
%frameworks from workshop organizers---Momeni's 
reproducibility checklists~\cite{momeni2025checklists}, Bleier and Khan's Methods Hub~\cite{bleier2025methodshub,khan2025valistad}, and Dess{\`i}'s knowledge graphs and TeleScope dataset~\cite{gangopadhyay2023claimskg,gangopadhyay2025telegram}, the research community can establish \emph{reproducible provenance} as the foundation for countering GenAI disinformation.

As one respondent concluded:

\begin{quote}
\emph{``We must treat information integrity as infrastructure. Just as we build roads and power grids, we must build the protocols for truth verification.''} \hfill (Policy Advisor)
\end{quote}

%This report submits that R2CASS is the forum where those 
Protocols must be defined, tested, and shared.

\paragraph{Ongoing Research.} This paper presents findings from Wave~1 of our longitudinal study. We invite domain experts---AI researchers, policymakers, journalists, and cybersecurity professionals---to participate in Wave~2. The survey is available at \url{https://github.com/aloth/verification-crisis}.

\balance
\bibliographystyle{ACM-Reference-Format}
\bibliography{bibliography}

\end{document}